\documentclass[]{aa}  

\usepackage{natbib}
\bibpunct{(}{)}{;}{a}{}{,} 
\usepackage{graphicx}
\usepackage{txfonts}
\begin{document}
   \title{Evaluating the stability of atmospheric lines with HARPS\thanks{Based on observations 
   taken at the 3.6m telescope at La Silla.
}}
   \author{P. Figueira, 
          F. Pepe,
          C. Lovis,
             \and 
	M. Mayor
          }

   \institute{Observatoire Astronomique de l'Universit\'{e} de Gen\`{e}ve, 51 Ch. des Maillettes 
    - Sauverny - CH1290, Versoix, Suisse\\
              \email{pedro.figueira@unige.ch}
          }

   \date{}

  \abstract{In the search for extrasolar systems by radial velocity technique, a precise wavelength calibration is necessary for high-precision measurements. The choice of the calibrator is a particularly important question in the infra-red domain, where the precision and exploits still fall behind the achievements of the optical.}{We investigate the long-term stability of atmospheric lines as a precise wavelength reference and analyze their sensitivity to different atmospheric and observing conditions.}{We use HARPS archive data on three bright stars, Tau\,Ceti, $\mu$ Arae and $\epsilon$\,Eri, spanning 6 years and containing high-cadence measurements over several nights. We cross-correlate this data with an O$_2$ mask and evaluate both radial velocity and bisector variations down to a photon noise of 1\,m/s.}{We find that the telluric lines in the three data-sets are stable down to 10\,m/s (r.m.s.) over the 6 years. We also show that the radial velocity variations can be accounted for by simple atmospheric models, yielding a final precision of 1-2\,m/s.}{The long-term stability of atmospheric lines was measured as being of 10\,m/s over six years, in spite of atmospheric phenomena. Atmospheric lines can be used as a wavelength reference for short-time-scales programs, yielding a precision of 5\,m/s "out-of-the box". A higher precision, down to 2\,m/s can be reached if the atmospheric phenomena are corrected for by the simple atmospheric model described, making it a very competitive method even on long time-scales.}


\authorrunning{P. Figueira et al.}
\titlerunning{The stability of atmospheric lines as seen by HARPS}

   \maketitle
%

\section{Introduction}


   The discovery of a hot Jupiter orbiting 51\,Peg by \citet{1995Natur.378..355M} by radial velocity (RV) measurements triggered the quest for extrasolar planets. This breakthrough was only made possible through the usage of very precise wavelength calibration systems. The Th-Ar emission lamp, used with the cross-correlation function (CCF) method  \citep{1996A&AS..119..373B}, and the I$_2$ cell, explored with the deconvolution procedure \citep{1996PASP..108..500B} were extensively used to find planets by RV. Recent technological developments allowed for more precise spectrographs to be built, such as HARPS \citep{2003Msngr.114...20M}, and reduction and analysis methods have been perfected through the years \citep{2007A&A...468.1115L}. As of today, HARPS yields the most precise RV measurements, with sub-m/s precision, allowing for a succession of ground-breaking detections of the lightest planets known \citep{2006Natur.441..305L, 2009A&A...493..639M}.  
   
   Given the proven stability of these two well-established wavelength references, little investigation was made on other viable alternatives. With time, RV  expanded into the infra-red (IR) domain, where wavelength calibration is still in its infancy and no method has established itself as the paradigm. 
   
In our attempt to measure RV with CRIRES at a very early stage of the instruments' life, we used atmospheric lines as wavelength reference \citep{2008A&A...489L...9H}. In a recent paper describing an improved data reduction \citep{2009arXiv0912.2643F} we reached a precision of 5\,m/s over a time scale of one week. This result was obtained in a RV standard star using CO$_2$ lines as wavelength reference. 

The usage of telluric lines as a precise wavelength reference goes back to the first attempts of precise RV measurements, by \cite{1973MNRAS.162..255G}, on Arcturus and Procyon. Ten years later, the studies of \cite{1982A&A...114..357B}, \cite{1982ApJ...253..727S}, and \cite{1985A&A...149..357C}, also explored the usage of atmospheric lines as a viable alternative for wavelength calibration.  Using O$_2$ lines, these authors showed back in the 80's that a precision of 5\,m/s was within reach. The value is made even more relevant by the fact that they used different RV determination methods (different instrumentation, different line fitting approaches, etc.).  Recently \cite{2004MNRAS.353L...1S} used the same principle on H$_2$O lines and reached a precision of 5-10\,m/s on UVES data. In light of these results, two questions follow:
   
\begin{itemize}

\item What is the stability of atmospheric lines over long time-scales?

\item What is the sensitivity of atmospheric lines to different observing conditions?

\end{itemize}   
   
   In order to answer these two questions, we turned to HARPS archive data, now spanning 6 years. The high internal stability of HARPS leads to an unequalled precision in RV measurements.  The RV variations of atmospheric lines can then be assessed against the very precise wavelength calibration provided by Th-Ar. In this paper we answer the two previous questions and conclude on the suitability of atmospheric lines as a wavelength anchor.
  
  The paper is structured as follows. In Sect. $2$ we describe HARPS and the data-sets used in our investigation. Section $3$ describes the principles of our method and data reduction. The results are presented in Sect. $4$ and discussed in Sect. $5$. We conclude in Sect. $6$ with the lessons to learn from this campaign.
   

\section{Observations}

\subsection{Using HARPS as ``absolute reference"}

HARPS \citep{2003Msngr.114...20M} is a fiber-fed cross-dispersed echelle spectrograph installed at the 3.6m telescope in La Silla. The main dispersion is provided by an R4 echelle grating in Littrow configuration. The orders are then dispersed in a direction perpendicular to the dispersion direction by a grism and imaged on a 2$\times$2k4k CCD mosaic. This optical design creates 72 orders that span the whole optical range form 3785 to 6915\,$\AA$. The spectral resolution was measured as being of R$=$110 000 and the mean dispersion as of 0.015\,$\AA$/pxl (820 m/s). The sampling is of 3.3 pixel per resolution element. The instrument is located in a vacuum vessel to avoid spectral drift due to temperature and air pressure effects, which are kept below 0.01\,K and 0.01\,mbar, respectively. A Th-Ar emission lamp is used for wavelength calibration.

The very high stability of HARPS allows for a precision of 1\,m/s to be reached routinely. When a precision better than 1\,m/s is required, a second channel can be used to image the Th-Ar simultaneously with the science target. HARPS proven intrinsic IP stability permits to study spectral lines profile 
variations as well, which can be done using the well-know bisector technique (BIS) on the Cross-Correlation Function (CCF). It is important to note that HARPS long-term precision was measured as being lower than 1\,m/s (see previous references), making this study possible for the first time down to this level of precision.

\subsection{The selection of targets}

In order to perform our study we started by selecting targets bright enough to allow us to reach a photon noise of $\sim$1\,m/s on the atmospheric lines position error in individual exposures. Out of these we selected stars which had a high number of points ($>$\,1000), covered a long time span and had, preferably,  a very high cadence of measurements. We payed particular attention to stars followed in asterosismology campaigns. These campaigns, aiming at studying the stellar RV variations on a timescale of hours to days, followed the star across the sky throughout several consecutive nights.

By applying the mentioned criteria we pinpointed Tau\,Ceti (HD\,10700), $\mu$\,Ara (HD\,160691)  and $\epsilon$ Eri (HD\,20794) as the most suitable targets for our tests. The details on the relevant properties of the data sets are provided in Tab.\ref{obs_table}. Tau\,Ceti and $\mu$\,Ara were followed in asterosismology campaigns, of 3 and 8 days, respectively.  It is important to note that the beginning of data acquisition on these targets goes back to 2003, ensuring a long time span. 

\begin{table*}
\centering
\begin{tabular}{lcccccc} \hline\hline
 \ \ Target & $\#$ of observations & $\#$ of days with observations &  $\overline{\mathrm{\# observations/day }}$ & time span [d]  & $\overline{\mathrm{S/N}}$ & \\ \hline \hline
Tau\,Ceti   & 5270   & 110 & 47.9 & 2308  & 260  \\
$\mu$\,Ara & 2868 & 117 &  24.5 & 2303 & 176 \\
$\epsilon$\,Eri & 1527  & 104 & 14.7 & 2217 & 316 \\
\hline \hline
\end{tabular}

\caption{The summary of the data set properties for the stars used in this paper. Note that the S/N is calculated at the center of order 60.}\label{obs_table}
\end{table*}

Our objective was simply to provide for a strong background against which the telluric lines can be defined at high S/N and in a short integration time. The stellar spectrum in itself has very little influence on the obtained RV; the only eventual issue is the blending between stellar and atmospheric lines.


\section{Methodology \& Data Reduction}

A dedicated pipeline was created for the reduction of HARPS spectra, named DRS\footnote{Standing for {\it Data Reduction Software}.}. For more details see \cite{2003Msngr.114...20M}. In a nutshell, the pipeline provides the typical bad pixel/column bias and dark correction, flat-fielding and optimal extraction of the spectra. The spectra are then calibrated in wavelength by using a Th-Ar lamp. The wavelength-calibrated spectra are cross-correlated \citep[as described in ][]{1996A&AS..119..373B} with a weighted template mask \citep{2002A&A...388..632P}. In order to calculate the RV of the atmospheric lines present in the spectra with the existing pipeline one needs then to build a telluric mask.

Using HITRAN database \citep{2005JQSRT..96..139R} one can locate 11992 spectral lines within HARPS wavelength domain. These lines correspond to H$_2$O, OH and O$_2$ molecules (with 8706, 3060, and 226 lines, respectively), and concentrate on the red side of the spectra, from 540 to 690\,nm. Water lines cover the widest wavelength domain, but are characterized by a very high dynamic range and are strongly blended. Most of the OH lines are too faint to be detected and are not even reported on DelbouilleÕs FTS spectrum. Out of the 226 Oxygen absorption features one can find 71 lines deeper than 1\% and some as deep as 95\%. These lines are distributed in 2 bands, located at 628-663 and at 687-690\,nm. The features are spectrally well-separated by HARPS, showing no clumping among themselves neither with stellar lines, unlike the other two species. The presented arguments led us to build a mask composed only of O$_2$ lines. The vacuum wavelengths were drawn from HITRAN database, and the corresponding values for standard temperature and pressure were calculated. 
The depth of each line could not be accurately estimated by fitting gaussian functions on the high S/N 
spectra; for deep lines ($>$\,90\%) the departure from gaussian profile was too pronounced and the errors incurring from this assumption non-negligible. As an alternative we used the Delbouille 
FTS spectra\footnote{The original reference being $http://cdsads.u-strasbg.fr/abs/1973apds.book.....D$, the data is made available also by the $BASS\,2000$ database at $http://bass2000.obspm.fr/solar_{}spect.php$ in a much easier way to use.}. This high-resolution spectra was convoluted with an instrumental profile representative of HARPSÕ resolution and the resulting spectra analyzed. The depth of each line can then be estimated simply by measuring the depth of the synthetic absorption line. The depth of the corresponding spectra was found to differ by only 1-2\, \% relative to the normalized observed spectra depth. The objective of this procedure was to assign the correct depth and thus the correct weight to each line when calculating the cross-correlation, as described in  \cite{2002A&A...388..632P}.

The pipeline delivers with the RV the photon noise error on it, estimated from the analysis of the CCF function using the method described in \cite{2001A&A...374..733B}. Along with each RV calculation, the bisector inverse slope was computed on the CCF function following the procedure described in \cite{2001A&A...379..279Q}. 


\section{Results \& Analysis}

We excluded from our analysis the spectra which yielded a very low photon noise accuracy, superior to 5\,m/s. These correspond to a very small fraction of the measurements: 0.8\,\% for Tau Ceti,  1.08\% for $\mu$\,Ara and 0.2\% for $\epsilon$\,Eri. 

\begin{figure}
\includegraphics[width=9cm]{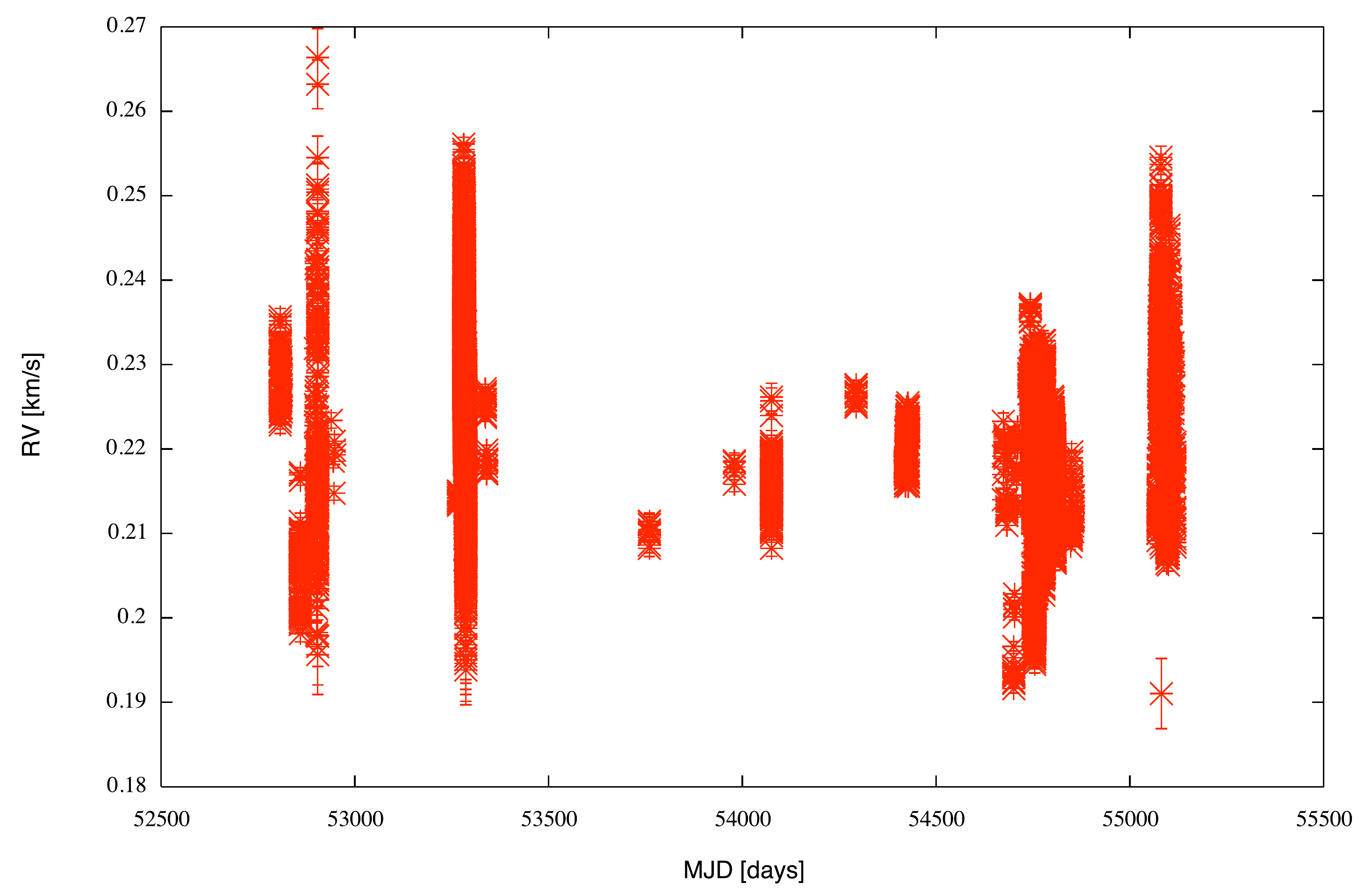}
\caption{The telluric RV variation for all Tau\,ceti measurements as function of time.}\label{RV_TauCeti}
\end{figure}

The scatter of the RV measurements and average photon noise for each star is presented in Tab. \ref{results}. It stands out that the RV variations are remarkably low, roughly of 10\,m/s for the three stars over 6 years.  We plot the RV as a function of time for Tau\,Ceti in Fig. \ref{RV_TauCeti}. This low scatter is a striking result, especially if we consider that there is no modeling, correction, or filtering involved. Still, the dispersion is much higher than the photon noise, showing that there is an additional source for the measured scatter.

\begin{table}
\centering
\begin{tabular}{lcc} \hline\hline
 \ \ Target & $\sigma$\,[m/s] & $\sigma_{ph}$\,[m/s]   \\ \hline \hline
Tau Ceti   & 10.74   & 0.98  \\
$\mu$\,Ara & 10.31 & 1.35 \\
$\epsilon$\,Eri & 10.82 & 0.76 \\

\hline \hline
\end{tabular}

\caption{The dispersion and average photon noise of the stars used in our campaign.}\label{results}
\end{table}

\begin{figure*}
\includegraphics[width=9cm]{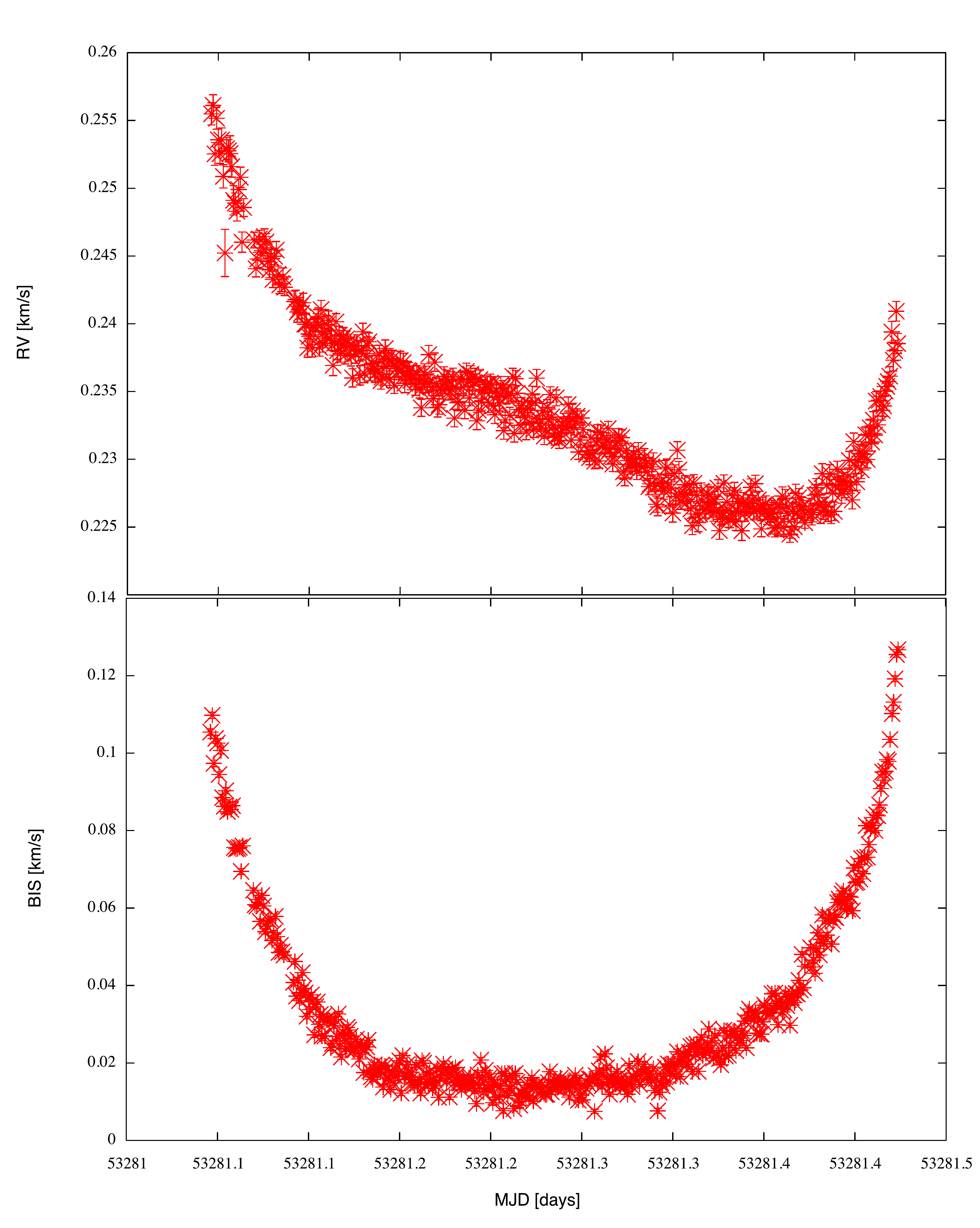}\includegraphics[width=9cm]{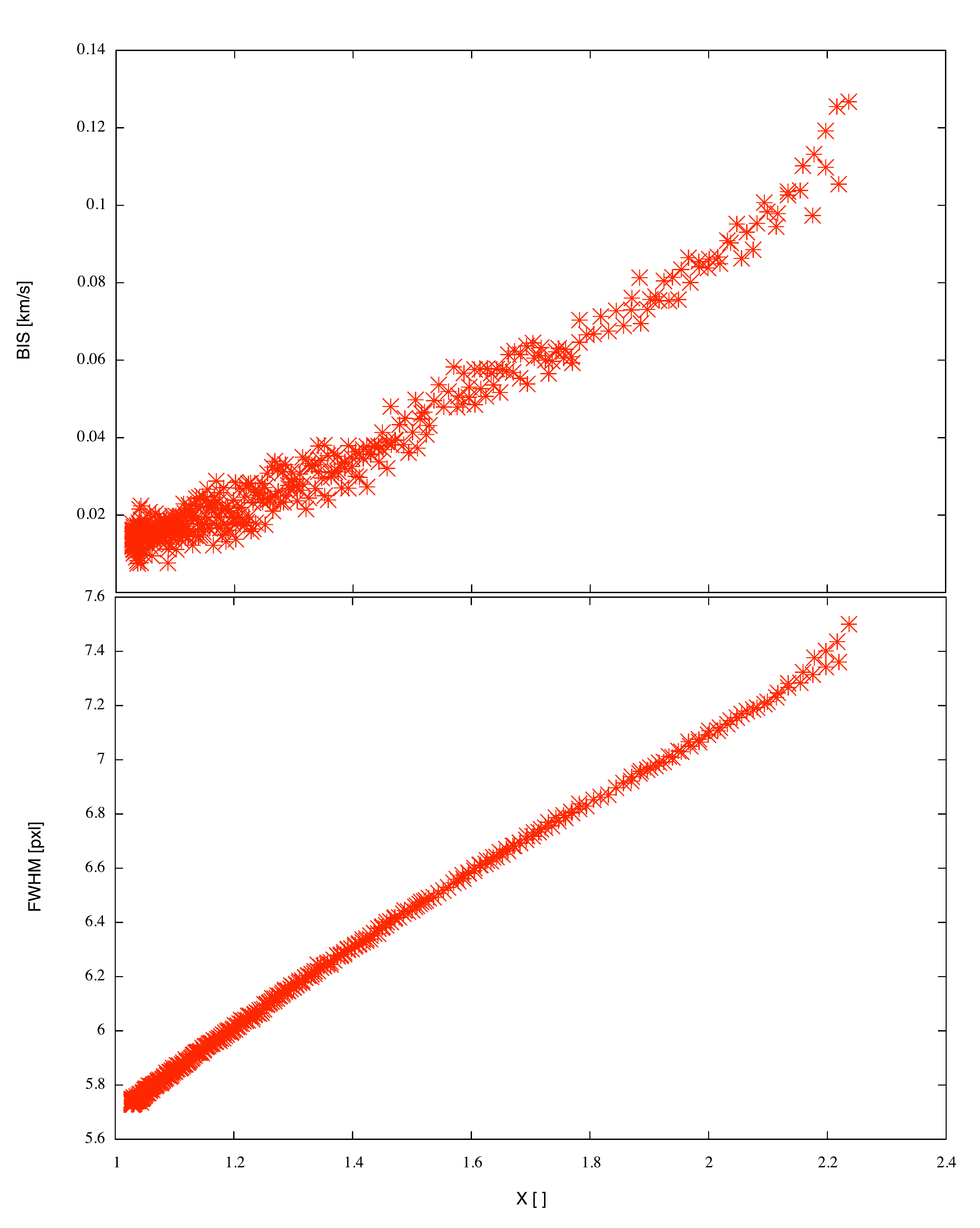}
\caption{Telluric RV measurements on Tau\,Ceti over a full night. Note the clear shape drawn by the RV {\it (left panel, top)} and the associated bisector {\it (left panel, bottom)} as function of time. In the right panel are depicted the correlation between BIS and airmass {\it (right panel, top)} and FWHM and airmass {\it (right panel, bottom)}. The the plotted error bars on RV and BIS correspond to photon errors. Photon errors on the BIS are approximated as being twice the RV errors.}\label{Tauceti_1n}
\end{figure*}

It is interesting to note that data subsets which cover a short time scale interval, such as one night, already exhibit an RV variation larger than the average photon noise. This points to a dependence of the measured RV on a factor which operates at a time scale of one night. In Fig. \ref{Tauceti_1n} we present the RV variation of atmospheric lines on the spectra of Tau\,Ceti over a full night ({\it upper left panel}), which illustrates very clearly this point.

The first thing to note in this series of measurements is the variation of BIS of the CCF throughout the night {\it (lower left panel)}. The BIS seems to be linearly correlated with the airmass {\it(upper right panel)}, and the FWHM of the fitted CCF as well {\it(lower right panel)}. Omitted from this series of plots is the correlation between contrast and airmass, very similar to the correlation between FWHM and airmass.
It stands out as very clear that observing a star at different airmass creates different absorption features, with a contrast and FWHM which are linearly dependent on airmass. On top of that, the line profile variation is not symmetric, as the BIS plot testifies, and scales with airmass as well. As a consequence, this asymmetry is expected to have an impact on the measured RV. 

The previously mentioned properties can be explained by the fact that some molecular lines are naturally asymmetric \citep[e.g.][]{2006CIAG.....F}\footnote{Note that this is the case for homonuclear diatomic molecules, like O$_2$.}. If one observes at different airmass the light passes through a medium with different path lengths. Because absorption does not vary linearly with geometric path length but exponentially, the line shape is bound to change. As one observes at higher airmass, the absorption introduced by the extra molecules will be stronger on the wings than in the core, leading to a larger, deeper and more asymmetric absorption line. Note that the CCF contrast oscillates around 72\%, and many lines have a contrast larger than 90\%. These lines are not in the regime linear anymore. Another possible explanation is that the line broadening as a function of pressure make the asymmetry detectable. This was measured for H$_2$ in the Solar System gaseous planets by \cite{1983ApJ...271..859C}.

However, and as seen on Fig. \ref{Tauceti_1n}, this effect alone cannot explain the shape drawn by the RV, and another effect that operates at a time scale of one night must be present. From our previous discussion on the airmass/bisector influence, it stems that this second effect does not have an impact on the line asymmetry; it must then be a phenomenon that shifts the center of atmospheric lines. 

One effect that presents these properties is the presence of a predominant horizontal wind, constant over one night. As one follows a star across the sky, the projection of the wind-vector along the line of sight of the telescope changes. For a given telescope elevation $\theta$ and azimuth $\phi$, the projection of an horizontal wind with direction $\delta$ and speed $w_s$ along the line of sight of the telescope is given by:

\begin{equation}
	w_{s,proj} =  w_s \times cos(\theta) \times cos(\phi -  \delta)
\end{equation}

Unfortunately, both these parameters are unknown\footnote{The weather stations at La Silla monitor the wind at atmospheric ground layer. It is well known that the wind at the ground layer is detached from those of high altitude, and as a consequence we cannot use weather monitor values as a proxy for neither $w_s$ nor $\delta$.}. Given the high number of points and good sampling of the parameters space, one can try to fit the mentioned effects. The measured RV is expected to be proportional to 1) airmass  and 2) wind velocity projected along the line of sight. This corresponds to determine $\alpha$, $\beta$, and $\delta$ such that the quantity

\begin{equation}
	\Omega =  \alpha \times  {\left({1\over sin(\theta)} - 1\right)} + \beta \times cos(\theta) \times cos(\phi -  \delta) + \gamma \label{eq_fit}
\end{equation}

 is linearly correlated with the measured RV (note that we used the fact that {\it airmass =  }$sin^{-1}(\theta)$). The $\gamma$ represents the zero-point of the RV. This quantity can be different from zero m/s, due to an error on the lines vacuum wavelength or on the wavelength conversion from vacuum to air.

We used a weighted linear-least squares minimization to determine the best fit parameters for each subset of data.  This approach was employed for subsets corresponding to complete nights, for which the hypothesis of a constant wind is a reasonable one. We compared the results yielded by linear least-squares with that of non-linear least-squares (using a Levenberg-Marquardt algorithm) and we recovered the same results.

If Eq. \ref{eq_fit} correctly describes the phenomena that affect the RVs, then, by construction, the parameter $\gamma$ is expected to be constant over time. From the same equation it stems that this parameter is simply the RV as measured at zenith, where the airmass and wind projection effects are null. One can then fit the different data-sets while requiring that the $\gamma$ is the same for all. By the same reasoning, $\alpha$ can be fixed. This is justified by the assumption that the bisector effect arises from the fundamental asymmetry of line and is thus not expected to vary with time. 

The results of the fitting are presented in Tab. \ref{funcstats}. We present the fitted parameters, along with  the dispersion before and after the fitted function is subtracted from the data, and the $\chi^2_{red}$ for each fit.

\begin{table*}
\centering
\begin{tabular}{lcccccccccc} \hline\hline
 \ \ Target & data set  &  $\#$obs  & $\sigma$\,[m/s] &  $\sigma_{(O-C)}$\,[m/s]  &  $\sigma_{ph}$\,[m/s] & $\chi^{2}_{red}$  & $\alpha$\,[m/s] &  $\beta$\,[m/s]  & $\gamma$\,[m/s] &  $\delta$\,[$^{o}$] \\ \hline \hline
Tau\,Ceti &  2004-10-03 & 437 &  6.40 &   1.67 & 0.64 & \dag  &  17.75 &  43.39 & 222.01 & -167.21 \\
 & 2004-10-04 & 438 &  7.98 &   1.33 &   0.65 & \dag  &  --- &  27.89 & --- & -154.15 \\
 & 2004-10-05 & 599 &   7.12 &   2.03 &   0.79 & \dag &  --- &  15.17 & --- & -133.95 \\
\hline

$\mu$\,Ara & 2004-06-04 & 278 &  6.90 &   1.90 &   1.27 & \dag &  --- &  33.27 & --- & -155.37 \\
& 2004-06-05 & 274 & 8.35 &   2.50 &   1.30 &  \dag &  --- &  29.34 & --- & -140.20 \\
& 2004-06-06 & 285 &  8.94 &   1.72 &   1.11 & \dag &  --- &  27.45 & --- & -136.20 \\
&  2004-06-07 & 286 &  4.48 &   1.60 &   1.03 & \dag &  --- &  23.62 & --- & -165.43 \\
&  2004-06-08 & 275 & 3.98 &   1.81 &   1.07 & \dag &  --- &  36.61 & --- & -168.70 \\
& 2004-06-09 & 214 &  6.88 &   4.02 &   1.34 & \dag &  --- &  41.89 & --- & -164.93 \\
& 2004-06-10 & 202 & 6.92 &   2.55 &   1.81 & \dag &  --- &  41.74 & --- & -142.11 \\
& 2004-06-11 & 273 &  8.41 &   3.51 &   2.07 & \dag  &  --- &  48.87 & --- & -155.55 \\
 \hline
Both stars & all data & 3562 &11.79 &   2.27 &   1.09 &   4.01 & & & & \\

\hline \hline
\end{tabular}

\caption{The fitted parameters and data properties, before and after the fitted model is subtracted from it. In this fit $\alpha$ and $\gamma$ are imposed to be the same for all data sets. Note that since the fit is made simultaneously for all data sets, the $\chi^2_{red}$ calculation is not applicable for a single night and the respective table entries are signaled by a \dag. The Table structure is left unchanged to allow for an easier comparison with Tab. \ref{funcstats_1} and Tab. \ref{funcstats_2}. Note that $\delta$=0 corresponds to the South-North direction.}\label{funcstats}
\end{table*}

\begin{figure}
\includegraphics[width=9cm]{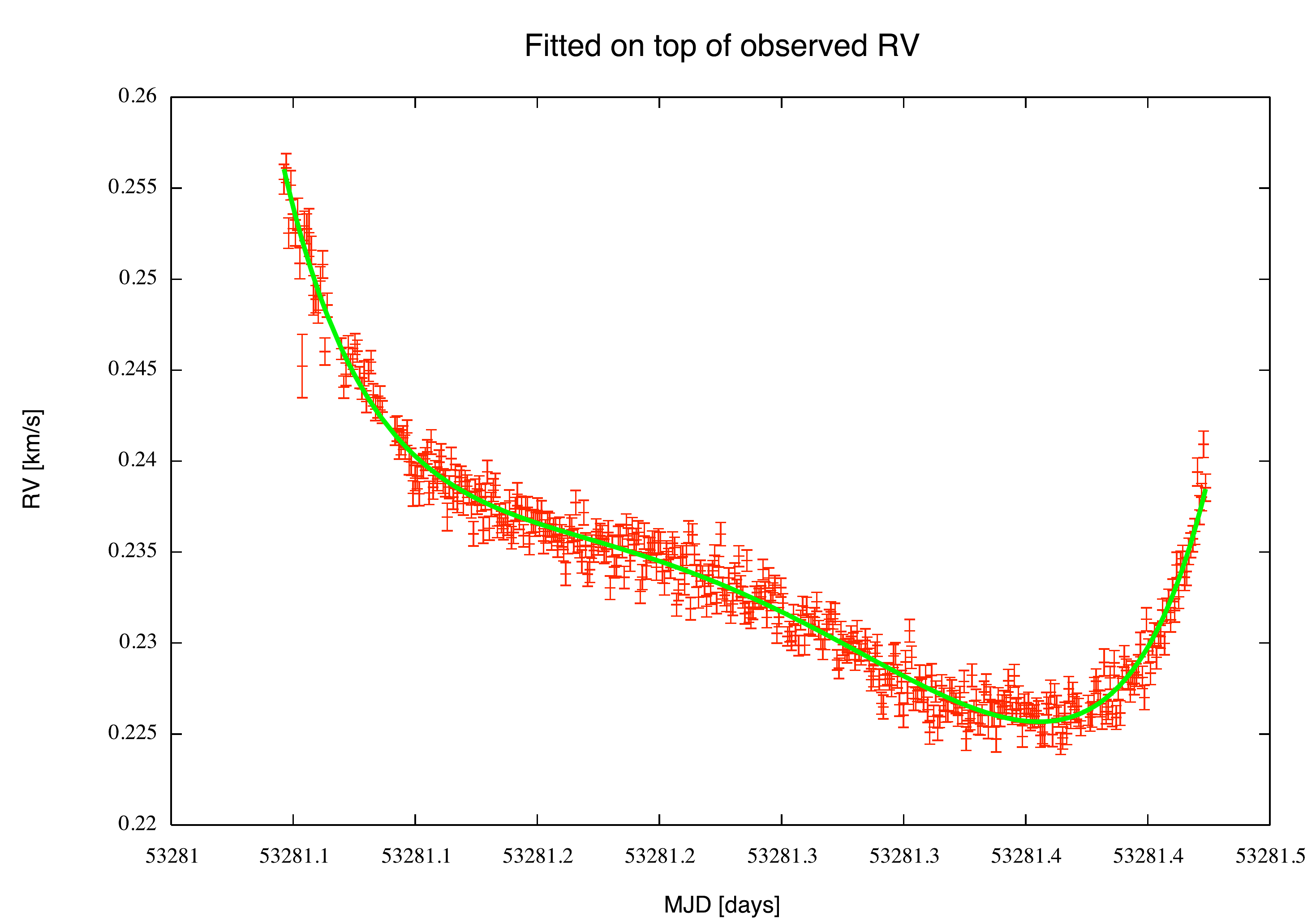}
\caption{The fit of atmospheric variation for the first night of the asterosismology run of Tau\,Ceti. The fitted model is described by Eq. \ref{eq_fit} and the parameters are presented in Tab \ref{funcstats_1}.}\label{Fit_Tauceti}
\end{figure}

We considered also the case in which $\alpha$ and both $\alpha$ and $\gamma$ are allowed to vary. The results are presented in the Appendix A in Tab. \ref{funcstats_1} and Tab. \ref{funcstats_2}, respectively.

The low chi-square, within the range 2.5\,--\,4.01, attests for the validity of the empirical description of the RV variation. The residuals around the fit are around twice the photon noise, showing that this correction is effective down to roughly 2\,m/s. As an example the fit of the data set corresponding to the first night of Tau\,Ceti campaign is presented in Fig. \ref{Fit_Tauceti}. 


\section{Discussion}

Our results show that O$_2$ telluric lines are stable down to 10\,m/s (r.m.s.) on a timescale of 6 years, in spite of atmospheric phenomena. These introduce variations at the 1-10\,m/s level and can be accounted for by simple atmospheric models, which were described in Sect.  4.

The properties of the fitted parameters are very insightful on their own. The imposition of a constant $\alpha$ and  $\gamma$ for the different data-sets lead to $\beta$ and $\delta$ values which vary more smoothly over time and are thus more likely to correspond to the physical ones. It is important to note that we would expect exactly the opposite if the model assumptions were wrong: as the number of parameters was reduced, the remaining ones would have to vary more to compensate for the observed variations. We conclude then that the description of the phenomena is correct. Still, we cannot discard that our model cannot be refined; It might be that the airmass effect depends on a second order on the wind speed, for instance. In any case, it already provides very good results down to twice the photon noise. Allowing all the fitting parameters to vary reduces the $\chi^2_{red}$, but not in a very significant way; we consider thus the approach of fixing the $\alpha$ and $\gamma$ the most adequate.

The fitting yields commonly $\beta$ bigger than $\alpha$, i.e., the effect of wind projection on RV is usually of higher magnitude than the airmass effect. It is interesting to notice that the wind is predominantly from between N and NE and of roughly 40\,m/s, which is a common value for high-altitude wind over La Silla.

The constance of the wind over short time-span and the fact that the two effects scale with airmass lead to a very important consequence on its own. It attests that telluric lines can be straightforwardly used as a precise wavelength reference over short time-scales if one observes at low airmass. In Tab. \ref{dispairmass} we present the dispersion of RV over timescales of 1-8 days at restricted airmass, quantifying this conclusion. It is then clear that a precision of 5\,m/s can be routinely obtained without using any atmospheric corrections if one observes in these conditions.

It is also important to recall that both O$_2$ and CO$_2$ are abundant species in the atmosphere, with a roughly constant v.m.r. up to 80\,km. This makes the molecules less sensitive to short time-scale weather variations and the assumptions of a constant horizontal wind a more realistic hypothesis. 

\begin{table}
\centering
\begin{tabular}{lccccc} \hline\hline
 \ \ Target & data-set [d]   &  $\sigma_{(X<1.5)}$  &  $\sigma_{(X<1.2)}$  & $\sigma_{(X<1.1)}$\,[m/s]    \\ \hline \hline
Tau\,Ceti & 1  & 4.53 &   3.38 &   2.41\\
 & 2  & 5.53 &   4.07 &   3.27 \\
& 3  & 5.81 &   4.55 &   3.70 \\
& Full & 9.77 &   9.16 &   8.43 \\
\hline

$\mu$\,Ara & 1 &   5.00 &   3.36 &   2.31 \\
& 2 &   6.18 &   4.14 &   2.61 \\
& 3 &   6.38 &   4.27 &   2.66 \\
& 4 &   5.67 &   3.86 &   2.39 \\
& 5 &   5.67 &   4.01 &   2.57 \\
& 6 &   6.18 &   4.71 &   3.24 \\
& 7 &   6.46 &   4.93 &   3.40 \\
& 8 &   6.94 &   5.42 &   3.91 \\

& Full &   9.88 &  10.10 &  10.23 \\

\hline

$\epsilon$\,Eri & Full & 10.82 &  10.55 &   9.61 \\

\hline \hline
\end{tabular}

\caption{The measured dispersion for data-sets with restricted airmass, for time-spans of 1-8 days and the whole data-set.}\label{dispairmass}
\end{table}

We did not check our mask for possible blending between telluric and stellar lines. The superposition of stellar lines on the telluric spectra is an effect which varies from star to star, due to the different mean stellar RV and the projection of Earth's motion around the Sun in the line of sight of the star. However, for this to bias the final RV, i.e. introduce a systematic error, the blends would have to be similar for all lines.  As a consequence the superposition between stellar and telluric lines will, if present, only lead to an increase in the scatter of the measured RVs. An extensive study is now starting in order to understand the impact of this effect.


\section{Conclusions}

Our results proves that O$_2$ telluric lines are stable down to 10\,m/s on a timescale of 6 years. Atmospheric phenomena introduce variations at the 1-10\,m/s level (r.m.s.). These can be accounted for by simple atmospheric models, which allow us to reach a precision of 2\,m/s, twice the photon noise.

Our results testify the power of atmospheric features as a wavelength calibrator. It is now clear that approach can compete with gas-cell method on short time-scales (of 1 week) in a straightforward way. In fact, it can even provide a better precision than gas cell at longer timescales if the atmospheric variations are characterized and compensated for by simple modeling.

\begin{acknowledgements}
     Support from the Funda\c{c}\~{a}o para Ci\^{e}ncia e a Tecnologia (Portugal) to P. F. in the form of a scholarship (reference SFRH/BD/21502/2005) is gratefully acknowledged. P.F. thanks to everyone who contributed to Planetary group meetings, and the referee Bill Cochran for his valuable comments on the paper.
\end{acknowledgements}

\bibliographystyle{aa} 
\bibliography{Mybibliog} 

\appendix

\section{Aditional fitting results}

In this section we present the parameters for the best fit of equation (\ref{eq_fit}), the dispersion of the data before and after the fit correction, the average photon noise and the $\chi^2_{red}$. Note that $\delta$=0 corresponds to the South-North direction.

\begin{table*}
\centering
\begin{tabular}{lcccccccccc} \hline\hline
 \ \ Target & dataset  &  $\#$obs  & $\sigma$\,[m/s] &  $\sigma_{(O-C)}$\,[m/s]  &  $\sigma_{ph}$\,[m/s] & $\chi^{2}_{red}$  & $\alpha$\,[m/s] &  $\beta$\,[m/s]  & $\gamma$\,[m/s] &  $\delta$\,[$^{o}$] \\ \hline \hline
Tau\,Ceti &  2004-10-03 & 437 &  6.40 &   1.05 &   0.64 &   2.52 &  32.34 &  83.91 & 212.53 & -173.24 \\
 & 2004-10-04 & 438 &  7.98 &   1.16 &   0.65 &   3.05 &  27.84 &  60.45 & 214.36 & -168.28 \\
 & 2004-10-05 & 599 &  7.12 &   1.40 &   0.79 &   3.33 &  32.75 &  75.65 & 208.01 & -171.96 \\
\hline

$\mu$\,Ara & 2004-06-04 & 278 &   6.90 &   1.75 &   1.27 &   1.92 &  24.31 &  64.66 & 235.34 & -167.43 \\
& 2004-06-05 & 274 &  8.35 &   2.13 &   1.30 &   2.15 &   1.39 &  27.65 & 207.39 & -43.82 \\
& 2004-06-06 & 285 &   8.94 &   1.69 &   1.11 &   2.15 &  23.61 &  43.38 & 229.04 & -154.23 \\
&  2004-06-07 & 286 &  4.48 &   1.58 &   1.03 &   2.39 &  24.81 &  46.31 & 230.45 & -172.64 \\
&  2004-06-08 & 275 &  3.98 &   1.63 &   1.07 &   2.35 &  32.92 & 100.02 & 246.72 & -175.53 \\
& 2004-06-09 & 214 &  6.88 &   3.36 &   1.34 &   2.63 &  34.04 & 121.93 & 252.77 & -172.50 \\
& 2004-06-10 & 202 &  6.92 &   2.36 &   1.81 &   1.74 &  -5.26 &  47.60 & 195.54 & -33.05 \\
& 2004-06-11 & 273 &  8.41 &   3.48 &   2.07 &   2.24 &  27.72 &  81.03 & 234.53 & -165.52 \\
 \hline
Both stars & all data & 3562 & 11.79 &   1.95 &   1.09 &   2.52 & & & & \\

\hline \hline
\end{tabular}

\caption{The fitted parameters and data properties, before and after the fitted model is subtracted from it, when all parameters are left to vary freely.}\label{funcstats_1}
\end{table*}

\begin{table*}
\centering
\begin{tabular}{lcccccccccc} \hline\hline
 \ \ Target & dataset  &  $\#$obs  & $\sigma$\,[m/s] &  $\sigma_{(O-C)}$\,[m/s]  &  $\sigma_{ph}$\,[m/s] & $\chi^{2}_{red}$  & $\alpha$\,[m/s] &  $\beta$\,[m/s]  & $\gamma$\,[m/s] &  $\delta$\,[$^{o}$] \\ \hline \hline
Tau\,Ceti &  2004-10-03 & 437 &  6.40 &   1.05 &   0.64 &  \dag &  31.62 &  81.42 & 213.09 & -173.03 \\
 & 2004-10-04 & 438 &  7.98 &   1.17 &   0.65 &  \dag &  29.44 &  66.03 & --- & -169.28 \\
 & 2004-10-05 & 599 &  7.12 &   1.47 &   0.79 &  \dag &  26.33 &  53.16 & --- & -168.52 \\
\hline

$\mu$\,Ara & 2004-06-04 & 278 &  6.90 &   1.90 &   1.27 &  \dag &   8.52 &  15.11 & --- & -106.45 \\
& 2004-06-05 & 274 & 8.35 &   2.15 &   1.30 &  \dag &   5.81 &  19.69 & --- & -76.20 \\
& 2004-06-06 & 285 &   8.94 &   1.78 &   1.11 &  \dag &  11.59 &  19.42 & --- & -78.89 \\
&  2004-06-07 & 286 &  4.48 &   1.67 &   1.03 &  \dag &  11.53 &   5.91 & --- & -83.57 \\
&  2004-06-08 & 275 &  3.98 &   1.86 &   1.07 &  \dag &   7.95 &  12.34 & --- & -142.69 \\
& 2004-06-09 & 214 &  6.88 &   3.91 &   1.34 &  \dag &   7.50 &  19.43 & --- & -139.67 \\
& 2004-06-10 & 202 & 6.92 &   2.41 &   1.81 &  \dag &   8.04 &  26.12 & --- & -107.04 \\
& 2004-06-11 & 273 &  8.41 &   3.59 &   2.07 &  \dag &  12.27 &  29.56 & --- & -136.75 \\
 \hline
Both stars & all data & 3562 &11.79 &   2.08 &   1.09 &  2.80 & & &  & \\

\hline \hline
\end{tabular}

\caption{The fitted parameters and data properties, before and after the fitted model is subtracted from it. In this fit $\gamma$ is imposed to be the same for all datasets. Note that since the fit is made simultaneously for all datasets, the $\chi^2_{red}$ calculation is not appliable for a single night and the respective table entries are signaled by a \dag. }\label{funcstats_2}
\end{table*}

\end{document}